\documentclass[%
 reprint,
 amsmath,amssymb,
 aps,
prb,
]{revtex4-1}

\usepackage{graphicx}% Include figure files
\usepackage{dcolumn}% Align table columns on decimal point
\usepackage{bm}% bold math
\usepackage{upgreek} % non-italic greek letters in math mode

\begin{document}

%\preprint{APS/123-QED}

\title{Simultaneous detection of the spin-Hall magnetoresistance and the spin-Seebeck effect in Platinum and Tantalum on Yttrium Iron Garnet}

%--------------------------------Author------------------------------------------------------------------------------------------------------------------
%-------------------------------------------------------------Author-------------------------------------------------------------------------------------
\author{N. Vlietstra}
% \email{n.vlietstra@rug.nl}
\affiliation{Physics of Nanodevices, Zernike Institute for Advanced Materials, University of Groningen, Groningen, The Netherlands}
%\\This line break forced with \textbackslash\textbackslash

\author{M. Isasa}
\affiliation{CIC nanoGUNE, 20018 Donostia-San Sebastian, Basque Country, Spain}
%\\This line break forced with \textbackslash\textbackslash

\author{J. Shan}
\affiliation{Physics of Nanodevices, Zernike Institute for Advanced Materials, University of Groningen, Groningen, The Netherlands}%\\This line break forced with \textbackslash\textbackslash

\author{F. Casanova}
\affiliation{CIC nanoGUNE, 20018 Donostia-San Sebastian, Basque Country, Spain}
\affiliation{IKERBASQUE, 48011 Bilbao, Basque Country, Spain}
%\\This line break forced with \textbackslash\textbackslash

\author{J. Ben Youssef}%
\affiliation{ 
Laboratoire de Magn\'etisme de Bretagne, CNRS, Universit\'e de Bretagne Occidentale, Brest, France%\\This line break forced with \textbackslash\textbackslash
}%

\author{B. J. van Wees}
\affiliation{Physics of Nanodevices, Zernike Institute for Advanced Materials, University of Groningen, Groningen, The Netherlands}%\\This line break forced with \textbackslash\textbackslash

\date{\today}% It is always \today, today,
             % but any date may be explicitly specified

\begin{abstract}

The spin-Seebeck effect (SSE) in platinum (Pt) and tantalum (Ta) on yttrium iron garnet (YIG) has been investigated by both externally heating the sample (using an on-chip Pt heater on top of the device) as well as by current-induced heating. For SSE measurements, external heating is the most common method to obtain clear signals. Here we show that also by current-induced heating it is possible to directly observe the SSE, separate from the also present spin-Hall magnetoresistance (SMR) signal, by using a lock-in detection technique. Using this measurement technique, the presence of additional 2$^\textrm{nd}$ order signals at low applied magnetic fields and high heating currents is revealed. These signals are caused by current-induced magnetic fields (Oersted fields) generated by the used AC-current, resulting in dynamic SMR signals. %For low applied magnetic fields, these current-induced magnetic fields will result in a deviation of the magnetization direction of the YIG, giving rise to 2$^\textrm{nd}$ order SMR signals.

\begin{description}

\item[PACS numbers]
72.25.Mk, 72.80.Sk, 75.70.Tj, 75.76.+j
\end{description}
\end{abstract}


\maketitle

%------------------------------------------------------------------------------------------------------------------------------------------------------------------
%------------------------------------------------------------------------------------------------------------------------------------------------------------------
%------Introduction------
\section{Introduction}
For the investigation of pure spin transport phenomena, yttrium iron garnet (YIG) is shown to be a very suitable candidate. YIG is a ferrimagnetic insulating material having a low magnetization damping as well as a very low coercive field. In combination with a high spin-orbit coupling material such as platinum (Pt), many different experiments have been performed, showing spin-pumping\citep{HillebrandsSPmagnons,SpinPump,SaitohFreqDep,CastelPRB}, spin transport\citep{travellingSW,Kajiwara2010nature} and spin-wave manipulation\citep{AzevedoAPL2013,Kurebayashi2011nmat,AzevedoPRL2011} as well as the recently discovered spin-Hall magnetoresistance (SMR).\citep{BauerTheorySMR,AlthammerSMR,VlietstraSMR,BauerSMR,VlietstraSMR2,JamalTaPt} 

Recently also experiments were performed showing the spin-Seebeck effect\citep{SSESaitoh,SSETheory,SSETheory2,reportSSE} (SSE) as well as the spin-Peltier effect\citep{PeltierYIG} in YIG/Pt systems. The SSE is observed when a temperature gradient is present over a ferromagnetic/non-magnetic interface. In a YIG/Pt system, this temperature gradient causes the creation of thermal magnons, resulting in transfer of angular momentum at the YIG/Pt interface, generating a pure spin-current into the Pt.\citep{reportSSE} This spin-current can then be detected electrically via the inverse spin-Hall effect (ISHE). So far, most experiments on the SSE are performed using external heating sources to create a temperature gradient over the device. Interestingly, Schreier et al.\citep{SSEGoennenwein} showed that a clear SSE signal can also be extracted from more easily performed current-induced heating experiments. In their experiments a temperature gradient is created by sending a charge current through the detection strip. A disadvantage of their measurement method is the presence of a much larger signal originated from the SMR, which should be subtracted to reveal the SSE signal.

%\footnote{The difference between the spin-dependent-Seebeck effect and the spin-Seebeck effect lies in the origin of the effect, where the first mentioned is a single electron effect, and the latter is caused by a collective motion of spins (spin waves/magnons)}\citep{Calori} 

In this paper we investigate both the SSE and SMR in a YIG-based device, showing the possibility to simultaneously, but separately, detect the SSE and SMR by using a lock-in detection technique. Whereas Schreier et al. only performed their measurements applying high magnetic fields, fully saturating the magnetization of the YIG, we show that when lowering the applied magnetic field, dynamic behavior of the magnetization of the YIG can be picked up as additional 2$^\textrm{nd}$ order signal. Only by using a lock-in detection technique these signals can be separately detected and analyzed. Having platinum (Pt) or tantalum (Ta) as detection layer, we investigate the evolution of the SSE and the SMR signal as a function of the magnitude and the direction of the applied field, focusing especially on their low-field behavior. 

The first experiments described in this paper show SSE measurements where a temperature gradient is generated by externally heating the sample using a second Pt strip on top of the device. By using devices consisting of both Pt and Ta on YIG, we confirm the opposite sign of the spin-Hall angle for Ta and Pt.\citep{SHE-Ta,SHE-Metals} In the secondly shown experiments, the samples are heated by current-induced heating through the metal detection strip, such that both the SSE and SMR are present. Additionally detected 2$^\textrm{nd}$ harmonic signals for low applied fields and high heating currents are discussed and ascribed to dynamic behavior of the magnetization of the YIG, caused by the applied AC-current. Finally, we derive a dynamic SMR term, which is used to explain the observed features.

The same kind of experiments could as well be used for detection of spin-transfer torque effects on the magnetization of the YIG, like the generation of spin-torque ferromagnetic resonance as formulated by Chiba et al.\citep{ChibaCurrent}. However, as will be shown in this paper, when applying low magnetic fields and high currents, the detected magnetization behavior is dominated by current-induced magnetic fields (like the Oersted field). So, to be able to detect any effect of the spin-transfer torque, its contribution should be increased, for example by decreasing the YIG thickness.

%------------------------------------------------------------------------------------------------------------------------------------------------------------------
%------------------------------------------------------------------------------------------------------------------------------------------------------------------
%------------------------------------------------------------------------------------------------------------------------------------------------------------------
%------Experimental details------

% Sample description
\section{Sample characteristics}
For the experiments shown in this paper, two Hall-bar shaped devices have been used, one consisting of a 5nm-thick Pt layer and the other of a 10nm-thick Ta layer. The Hall-bars have a length of 500$\upmu$m and a width of 50$\upmu$m, with side contacts of 10$\upmu$m width. Both Hall-bars are deposited on top of a 4x4mm$^2$ YIG sample, by dc sputtering.

The used sample consists of a 200nm thick layer of YIG, grown by liquid phase epitaxy on a single crystal (111)Gd$_3$Ga$_4$O$_{12}$ (GGG) substrate. The YIG magnetization shows isotropic behavior of the magnetization in the film plane, with a low coercive field of only 0.06mT.\citep{CastelPRB,VlietstraSMR} %Two Hall-bars, one consisting of a 5nm thick Pt layer and the other of a 10nm thick Ta layer, are deposited on top of the YIG by dc sputtering. %The thickness of the deposited layers was measured by atomic force microscopy with an accuracy of $\pm$0.5nm. 

For external heating experiments, a Ti/Pt bar of 5/40nm thick is deposited on top of both Hall-bars, separated from the main channel by a 80nm-thick insulating Al$_2$O$_3$ layer. The size of the heater is 400x25$\upmu$m$^2$. Finally, both Hall-bars and Pt heaters are contacted by thick Ti/Au pads [5/150nm]. All structures are patterned using electron-beam lithography. Before each fabrication step the sample has been cleaned by rinsing it in acetone, no further surface treatment has been carried out. A microscope image of the device is shown in Fig. \ref{fig:Fig1}(a).

\section{Measurement methods}
To observe the SSE, two measurement methods have been investigated. At first, to generate a clear SSE signal, a temperature gradient is created using an external heating source to heat one side of the sample.
% possible to add examples of heating plus some references
In our case, we have a Ti/Pt strip on top of the Hall-bar, electrically insulated from the detection channel, which can be used as an external heater. By sending a large current (up to 10mA) through the heater, the strip will be heated by Joule heating. Hereby, a temperature gradient will be formed over the YIG/Pt(Ta) stack, giving rise to the SSE.

A second method to generate the SSE, is the generation of a temperature gradient by current-induced heating through the detection strip. In this case a charge current is sent through the Hall-bar itself, which also leads to Joule heating, resulting in a temperature gradient over the YIG/Pt(Ta) stack. As the Hall-bar is directly in contact with the YIG, also the SMR will be present when using this heating method. 

To separately detect the SSE and the SMR signals, a lock-in detection technique is used. Using up to three Stanford SR-830 Lock-in amplifiers, the 1$^\textrm{st}$, 2$^\textrm{nd}$ and higher harmonic voltage responses of the system are separately measured. As SMR scales linearly with the applied current, its contribution will be picked up as a 1$^\textrm{st}$ harmonic signal. Similarly, the SSE scales quadratically with current, so its contribution will be detected as a 2$^\textrm{nd}$ harmonic signal. For lock-in detection an AC-current is used with a frequency of 17Hz. The magnitude of the applied AC-currents is defined by their rms values. %The advantage of this measurement technique is the possibility to obtain direct information about the scaling of the detected signals with the applied current. 

\begin{figure}
\includegraphics[width=8.5cm]{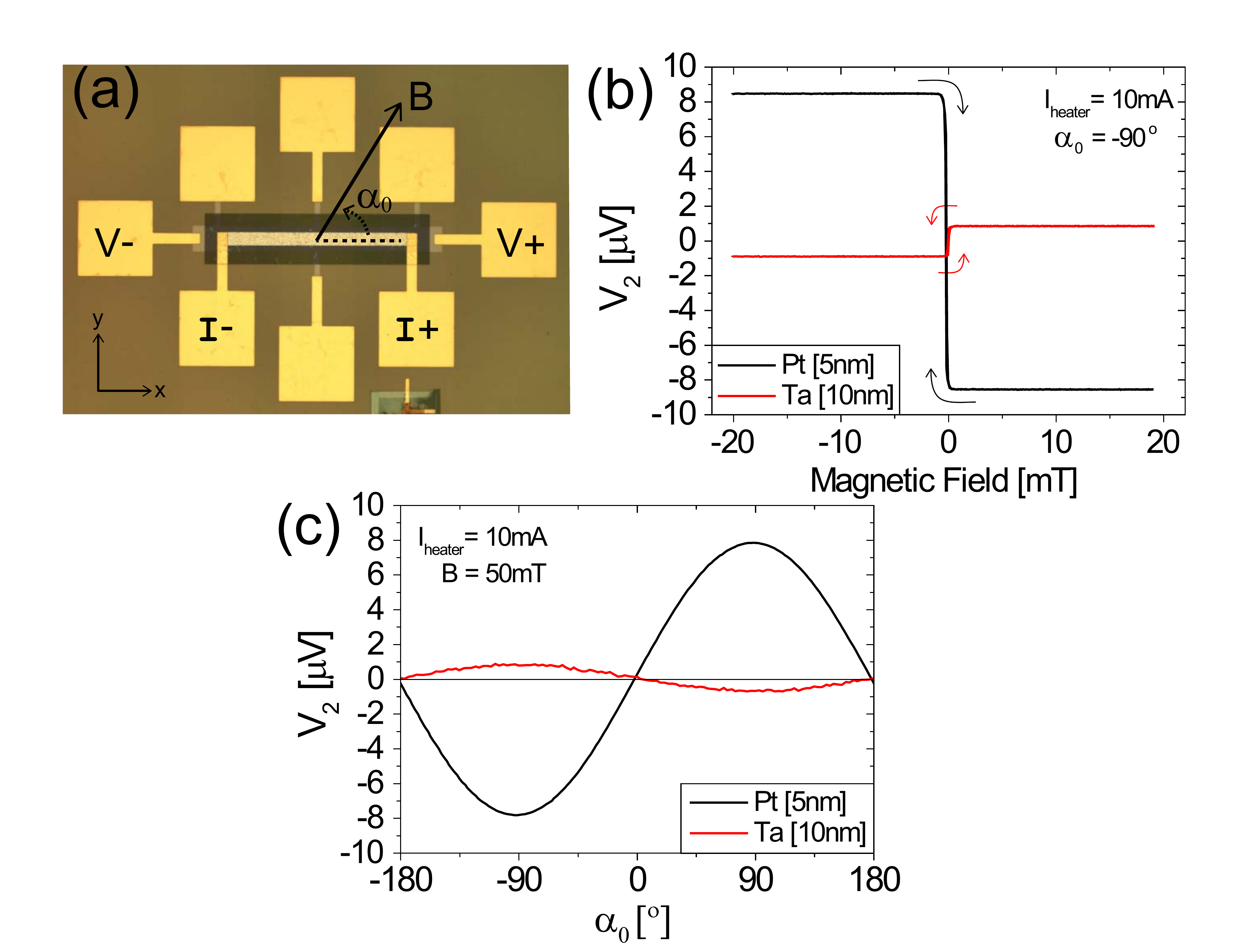}% Here is how to import EPS art
\caption{\label{fig:Fig1} 
(a) Microscope image of the device structure, consisting of a Pt or Ta Hall-bar detector (bottom layer) and a Pt heater (top layer), separated by an insulating Al$_2$O$_3$ layer. Ti/Au pads are used for contacting the device. For external heating experiments, the device is contacted as marked. The applied field direction is given by $\alpha_0$, as defined in the figure. (b) 2$^\textrm{nd}$ harmonic voltage signal generated by the SSE for a fixed magnetic field direction of $\alpha_0=90^\circ$ and (c) angular dependence of the SSE signal applying a magnetic field of 50mT, in both Pt and Ta.
}
\end{figure}

\begin{figure*}
\includegraphics[width=18cm]{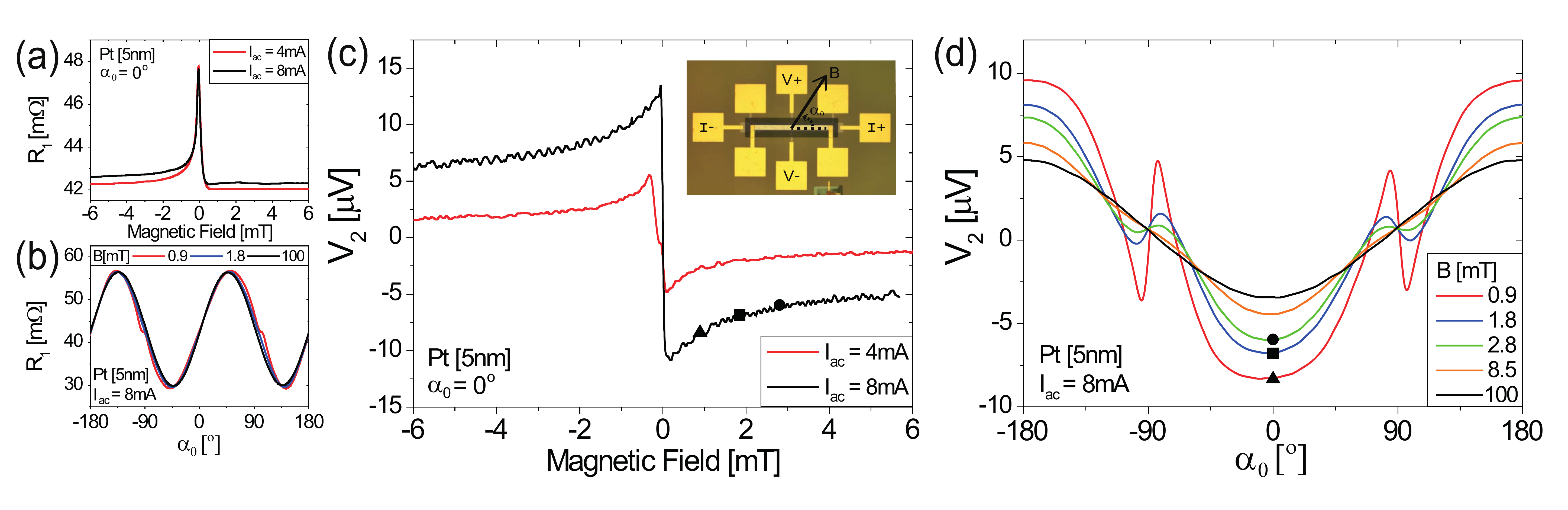}% Here is how to import EPS art
\caption{\label{fig:Fig2} 
Current-induced heating experiments on the YIG/Pt sample. (a) Magnetic field dependence of the 1$^\textrm{st}$ order resistance for $\alpha_0=0^\circ$, showing the SMR signal for applied AC-currents of 4mA and 8mA. (b) Angular dependence of the 1$^\textrm{st}$ order SMR signal for I$_\textrm{ac}$=8mA, for applied magnetic fields of 0.9mT, 1.8mT and 100mT. (c) and (d) show the corresponding 2$^\textrm{nd}$ harmonic voltage signals, respectively. For applied magnetic fields above 10mT, the 2$^\textrm{nd}$ harmonic response only shows the SSE signal. For low applied magnetic fields an additional signal is observed on top of the SSE signal. The black symbols in both figures are a guide for the eye. They show equal measurement conditions comparing the results shown in both figures. The inset of (c) shows the used measurement configuration. 
}
\end{figure*}

Evaluating the working mechanism of the lock-in detection technique in more detail (see appendix), shows that in order to obtain the linear response signal of the system, both the measured 1$^\textrm{st}$ harmonic signal as well as the 3$^\textrm{rd}$ harmonic signal have to be taken into account. Including both harmonic signals following the analysis explained in the appendix, we find the shown 1$^\textrm{st}$ order response. Note that the measured 2$^\textrm{nd}$ harmonic signal is directly plotted, without any corrections. 

In all experiments, an external magnetic field is applied to define the direction of the magnetization M of the YIG. The direction of the applied field is defined by $\alpha_0$, which is the in-plane angle between the current direction (along $x$) and the applied field direction, as it is marked in Fig. \ref{fig:Fig1}(a). Not only experiments at high saturation magnetic fields are performed, also the low field behavior is investigated. The applied magnetic field strength was measured by a LakeShore Gaussmeter (model 421) using a transverse Hall probe, to correct the set magnetic field for any present remnant field. All measurements are carried out at room temperature.

%------------------------------------------------------------------------------------------------------------------------------------------------------------------
%------------------------------------------------------------------------------------------------------------------------------------------------------------------
%------------------------------------------------------------------------------------------------------------------------------------------------------------------
%------Results and discussion------
\section{Results and Discussion}

\subsection{Spin-Seebeck effect by external heating}
For the external heating experiment an AC-current is sent through the top Pt strip as marked in Fig. \ref{fig:Fig1}(a). By measuring the 2$^\textrm{nd}$ harmonic voltage signals along the Hall-bar, the SSE is detected via the ISHE in Pt and Ta. Fig. \ref{fig:Fig1}(b) shows the typical SSE signals for both YIG/Pt and YIG/Ta samples for an applied field perpendicular to the longitudinal direction of the Hall-bar ($\alpha_0=90^\circ$). Changing the sign of B (and thus M) changes the sign of the signal, as the spin-polarization direction of the pumped spin-current is reversed. Due to the low coercive field of YIG almost no hysteresis is observed for the reversed field sweep. For the YIG/Pt and YIG/Ta sample opposite magnetic field dependence is observed, proving the opposite sign of the spin-Hall angle for Pt versus Ta. 

As the spin-polarization direction of the generated spin-current is dependent on the direction of the YIG magnetization, the SSE/ISHE voltage shows a sine shaped angular dependence with a period of 360$^\circ$. By rotating the sample in a constant applied magnetic field of 50mT, this angular dependence is detected as is shown in Fig. \ref{fig:Fig1}(c). Also here the effect of the opposite sign of the spin-Hall angle, for Ta compared to Pt, is clearly visible.  

From Fig. \ref{fig:Fig1} it is observed that the SSE signal for the YIG/Ta sample is almost a factor 10 smaller than for the YIG/Pt sample ($V_{SSE,Pt}/V_{SSE,Ta}=-9.8$). To compare, we calculate the expected ratio from the theoretical description of the SSE voltage, as reported by Schreier et al.\citep{SSETheory}:
\begin{equation}
	\label{eq:SSE}
	V_{SSE}=C_{YIG} \cdot \Delta T_{me} G_r \Theta_{SH} \rho l \eta \frac{\lambda}{t} \tanh{\left( \frac{t}{2\lambda}\right) }
\end{equation}
where $C_{YIG}$ contains all parameters describing properties of YIG, including some physical constants (defined in ref.\citep{SSETheory}), so $C_{YIG}$ is constant for both the YIG/Pt and the YIG/Ta sample. $\Delta T_{me}$ is the temperature difference between the magnons and electrons at the YIG/metal interface. $\rho$, $\lambda$, $t$ and $l$ are the resistivity, spin-diffusion length, thickness of the normal metal layer (Pt/Ta) and the distance between the voltage contacts, respectively. $\eta$ is the backflow correction factor, defined as
\begin{equation}
	\label{eq:backflow}
	\eta = \left[ 1 + G_r \rho \lambda \coth{\left(\frac{t}{\lambda}\right)} \right]^{-1}
\end{equation}

Previously, in ref.\citep{VlietstraSMR2}, we have determined the real part of the spin-mixing conductance at the YIG/Pt interface ($G_r=4.4\times10^{14} \Omega^{-1} \textrm{m}^{-2}$), the spin-Hall angle ($\Theta_{SH,Pt}=0.08$) and the spin-diffusion length ($\lambda_{Pt}=1.2$nm) of Pt. For the YIG/Ta sample we take the magnitude of these system parameters as reported by Hahn et al.\citep{JamalTaPt} ($G_r=2\times10^{13} \Omega^{-1} \textrm{m}^{-2}$, $\Theta_{SH,Ta}=-0.02$ and $\lambda_{Ta}=1.8$nm). As a check, we also used these parameter-values to calculate the 1$^\textrm{st}$ order SMR signals for Ta, and found good agreement with the measured signals (not shown). 

To get an estimate for $V_{SSE,Pt}/V_{SSE,Ta}$ we assume $\Delta T_{me}$ to be constant for both samples. By inserting the values of the mentioned parameters, the dimensions of the Hall-bars and the measured resistivity of the Pt and Ta layers ($\rho_{Pt}=3.4\times10^{-7}\Omega$m and $\rho_{Ta}=3.5\times 10^{-6} \Omega$m, respectively), we find $V_{SSE,Pt}/V_{SSE,Ta}=-10.6$, which is close to the experimentally observed ratio.

\subsection{Current-induced spin-Seebeck effect}
The second method used to detect the SSE is by current-induced heating through the metal detection strip itself, as recently was reported by Schreier et al.\citep{SSEGoennenwein}. In this section we show that we can achieve more directly similar results, by using a lock-in detection technique. By this technique, the SSE signals can directly be detected as a 2$^\textrm{nd}$ harmonic signal, fully separated from the SMR signal, which shows up in the 1$^\textrm{st}$ harmonic response. Furthermore, the lock-in detection technique enables us to reveal and investigate additional signals appearing when applying low magnetic fields.

The inset of Fig. \ref{fig:Fig2}(c) shows a microscope image of the sample, marking the position of the current and voltage probes for the current-induced heating experiments. The magnetic field direction is again defined by $\alpha_0$. This measurement configuration is similar to the method used to detect transverse SMR\citep{VlietstraSMR,VlietstraSMR2} and therefore we expect to observe SMR in the 1$^\textrm{st}$ order signal, as is shown in Figs. \ref{fig:Fig2}(a) and (b). In Fig. \ref{fig:Fig2}(b) it is observed that down to very low applied magnetic fields, the average magnetization direction of the YIG nicely follows the applied field direction, resulting in the $\textrm{sin}(2\alpha_0)$ angular dependence of the SMR.\citep{AlthammerSMR} Only for the lowest applied field of 0.9mT a small deviation of the signal around $\alpha_0=\pm90^\circ$ is observed, showing this field strength is not sufficient to assume M being (on average) fully along the applied field direction. %In Fig. \ref{fig:Fig2}(b) the SMR signal is visible as a peak around the switching field of the YIG magnetization. This peak shows the in-plane rotation of M towards the applied field direction.\citep{VlietstraSMR} SMR is more clearly shown in the angle-dependent measurement as shown in Fig. \ref{fig:Fig2}(d)

Similar to the external heating experiment, the SSE signal shows up in the 2$^\textrm{nd}$ harmonic signal. Figs. \ref{fig:Fig2}(c) and (d) show the magnetic field dependence and angular dependence of the detected 2$^\textrm{nd}$ harmonic signal, respectively. Comparing the shape of the 2$^\textrm{nd}$ harmonic data of the external heating experiments (Fig. \ref{fig:Fig1}(b)) to the current-induced heating experiments (Fig. \ref{fig:Fig2}(c)), an enhanced signal is observed in Fig. \ref{fig:Fig2}(c) for fields of a few mT. This additional signal cannot be explained by the angular dependence of the SSE, neither by the rotation of M in the plane towards B (by which the 1$^\textrm{st}$ harmonic SMR peaks in Fig. \ref{fig:Fig2}(a) are explained\citep{VlietstraSMR}). 

The angular dependence of this additional signal, as presented in Fig. \ref{fig:Fig2}(d), shows that besides an increased amplitude of the SSE signal (black symbols in Figs. \ref{fig:Fig2}(c) and (d)), also at $\alpha_0=\pm90^\circ$ additional peaks appear for low applied fields. By increasing the applied magnetic field, all extra signals disappear, leaving the expected SSE signal showing a 360$^\circ$ periodic angular dependence. %The magnitude of this signal scales (as expected) quadratically with current, so for low AC-currents (used for SMR experiments), the SSE will not be detected.  

\begin{figure}
\includegraphics[width=8.5cm]{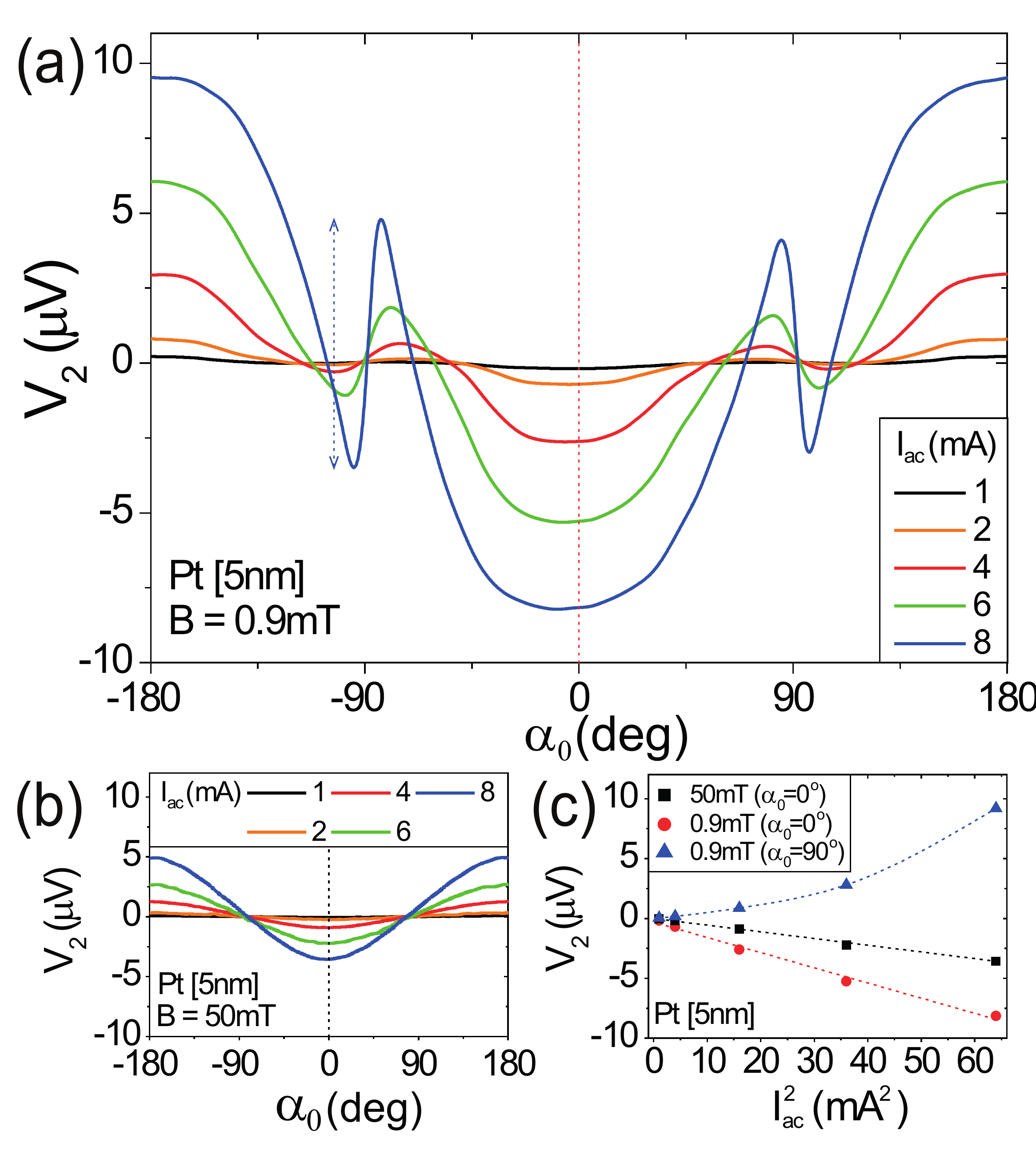}% Here is how to import EPS art
\caption{\label{fig:Fig3} 
AC-current dependence of the 2$^\textrm{nd}$ harmonic voltage for an applied field of (a) 0.9mT and (b) 50mT. For the shown experiments, the transverse current-induced heating measurement configuration has been used. The vertical dashed lines mark the data plotted in (c), which shows the AC-current dependence of the magnitude of the signal at $\alpha_0=0^\circ$ for B=0.9mT (red dots) and B=50mT (black squares) and the average magnitude of the peaks (peak to peak) around $\alpha_0=\pm90^\circ$ for B=0.9mT (blue triangles). The dashed lines are a guide for the eye. 
}
\end{figure}

To further characterize the additionally observed features at low applied magnetic fields, also their AC-current dependence has been measured and these results are shown in Fig. \ref{fig:Fig3}. It can be seen that the current dependence is very similar to the shown magnetic field dependence, giving maximal additional signals for low applied fields and high applied AC-currents. From Figs. \ref{fig:Fig3}(a) and (b), the magnitude of the signal at $\alpha_0=0^\circ$ is extracted and plotted separately in Fig. \ref{fig:Fig3}(c). As can be seen from this figure, for both the applied magnetic field of 0.9mT (Fig. \ref{fig:Fig3}(a)) and 50mT (Fig. \ref{fig:Fig3}(b)), the amplitude of the signal quadratically scales with the applied AC-current. The magnitude of the peaks around $\alpha_0=\pm90^\circ$, plotted in blue in Fig. \ref{fig:Fig3}(c), increases faster than quadratically, pointing to the presence of higher order effects.

%Following the theory of the SSE, its signal should not be dependent on the applied field strength, neither should overshoot the maximum value expected for the optimal angle of $\alpha_0=0^\circ$ or $180^\circ$.\citep{SSEGoennenwein} As the detected signals at low applied fields show both of these properties, another 2$^\textrm{nd}$ order effect should be present.

\begin{figure}[b]
\includegraphics[width=8.5cm]{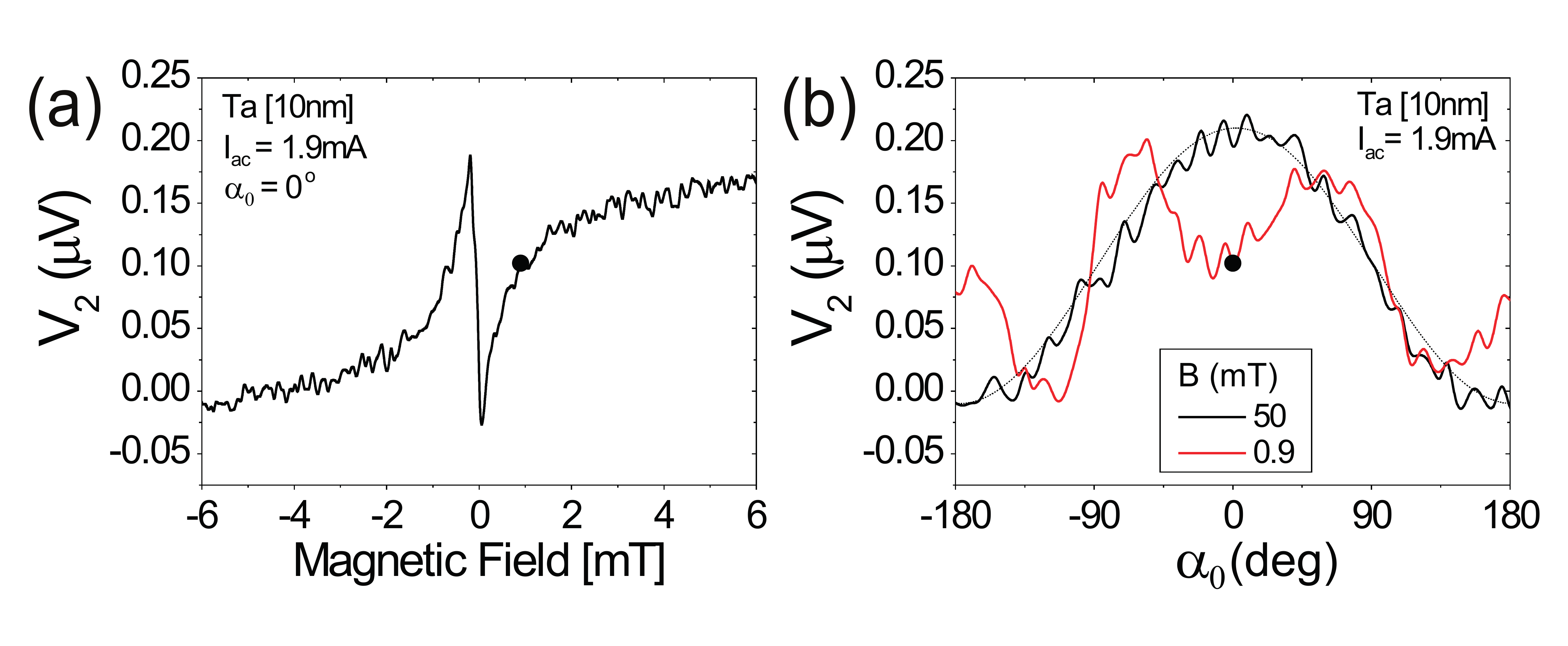}% Here is how to import EPS art
\caption{\label{fig:Fig4} 
2$^\textrm{nd}$ harmonic response of the transverse voltage for the YIG/Ta sample. (a) Magnetic field sweep for $\alpha_0=0^\circ$ and (b) angular dependence for two different applied magnetic fields (0.9mT and 50mT). The SSE signal (=signal at high field) has opposite sign compared to the YIG/Pt sample, whereas the low-field behavior is similar for both samples. The black symbols in the figures point out the equal measurement conditions comparing both figures.
}
\end{figure}

To fully exclude the SSE being the origin of the additionally detected signals, the current-induced heating measurements were repeated on the YIG/Ta sample. Results of those measurements are shown in Fig. \ref{fig:Fig4}. The applied current in those experiments is only 1.9mA, limited by the high resistance of the Ta bar ($\rho_{Ta}=3.5\times 10^{-6} \Omega \textrm{m}$). In both Fig. \ref{fig:Fig4}(a) and (b) it can be seen that the high-field signal nicely changes sign compared to the YIG/Pt data, as predicted for the SSE/ISHE, because of the opposite sign of the spin-Hall angle of Ta compared to Pt. Contrary, the low-field peak in the magnetic field sweep (Fig. \ref{fig:Fig4}(a)) keeps the same sign as in the YIG/Pt sample (Fig. \ref{fig:Fig2}(c)), showing the SSE cannot be the origin of this phenomenon. Furthermore, this result also excludes the observed feature being originated by any other effect linearly related to the spin-Hall angle of a material. So, possible deviations of M caused by spin-transfer torque, due to a spin-current created via the SHE, cannot directly be used to explain the observed features. Note that the SMR signal depends quadratically on the spin-Hall angle,\citep{BauerTheorySMR,VlietstraSMR2} which makes any effect related to the SMR a likely candidate for explaining the observed features.

Summarizing, the current-induced heating experiments show that when applying a sufficiently high magnetic field ($>$10mT), the SMR and SSE can be simultaneously, but separately, detected using an AC-current combined with a lock-in detection technique. By this method the SSE can thus be very easily and directly detected, without being interfered with the SMR signal. Furthermore, it is observed that for low magnetic fields, and/or high heating currents, additional signals appear on top of the SSE. The origin of these additional signals might be related to the SMR-effect, which will be discussed in more detail in the next section. 

%After passing the switching field of the YIG, the full curve changes sign, as expected from both the SSE and the observed behavior for the magnetic field sweep (not shown).

\subsection{Dynamic Spin-Hall Magnetoresistance}
%As described in the previous section, for applied magnetic fields above 10mT, the 2$^\textrm{nd}$ harmonic signal exactly shows the expected angular dependence for the SSE, which is independent of the magnitude of the applied field. However, by decreasing the applied field towards the switching field of the YIG, we observe a change in the angular dependence of the 2$^\textrm{nd}$ harmonic signal, showing an increasing amplitude, as well as additional peaks around $\alpha_0=\pm90^\circ$ (see Fig. \ref{fig:Fig2}(d)). 

During the measurements, large AC-currents are sent through the Hall-bar structure, which can generate magnetic fields. One source of these magnetic fields will be Oersted fields (B$_{oe}$) generated around the Hall-bar, perpendicular to the current direction. As the Hall-bar structure is very thin compared to its lateral dimensions, the Oersted field above/below the center of the bar can be estimated using the infinite plane approximation: $B_{oe}=\frac{\mu_0I}{2w}$, where $\mu_0$ is the permeability in vacuum, $I$ is the applied current and $w$ is the width of the Pt bar. Note that the generated field in this case is independent of the distance from the plane, so the full thickness of the YIG below the Hall-bar will be exposed to this field.\footnote{This approximation was checked by modeling the system using COMSOL Multiphysics, showing indeed a nearly constant Oersted field at least several hundreds of nanometers above/below the plane} For example, for an applied current of 8mA, an Oersted field of 0.1mT will be generated, which is significant compared to an applied magnetic field of 0.9mT. Therefore, for low applied fields, the magnetization direction of the YIG will also be affected by the generated Oersted field. 

As we are dealing with AC-currents, the generated Oersted field (and any other current-induced magnetic field) will continuously change sign, which might cause M to oscillate around the applied field direction.
%By applying large AC-currents through the Hall-bar structure, several effects can be present which might influence the magnetization direction of the YIG. First of all, Oersted fields are created, resulting in additional magnetic fields acting on M. Furthermore, other spin-orbit torques might influence M, such as Rashba or Dresselhaus fields.\citep{...}
In this case, $\alpha$ (defining the direction of M) is current-dependent,\citep{SOTnat} giving rise to dynamic equations for both the SMR and the SSE. The current-dependent behavior of the SMR signal is derived starting from the equation for transverse SMR\citep{BauerTheorySMR,VlietstraSMR2}
\begin{equation}
	\label{eq:SMR}
	V_{T,SMR}=IR_{T,SMR}=\Delta R_1Im_xm_y+\Delta R_2Im_z %=\Delta R_1sin(\alpha)cos(\alpha)+\Delta R_2sin(\beta)
\end{equation}
where $\Delta R_1$ and $\Delta R_2$ are resistance changes dependent on the spin-diffusion length, spin-Hall angle and spin-mixing conductance of the system, as defined in refs.\citep{BauerTheorySMR,VlietstraSMR2}. $m_{x,y,z}$ are the components of M pointing in respectively the x-, y-, and z-direction (where z is the out-of-plane direction). $m_x$ and $m_y$ can be expressed as $\sin(\alpha)$ and $\cos(\alpha)$, respectively. As the applied magnetic field is in-plane, as well as the generated Oersted field, combined with the large demagnetization field of YIG for out-of-plane directions, $m_z$ will be small and therefore neglected in further derivations. 

Small oscillations of M, due to the presence of AC-current generated magnetic fields, result in a current-dependent SMR signal, which can be expressed in first order as
\begin{equation}
	\label{eq:1storder}
	V_{T,SMR}(I) \approx V_{T,SMR} (\alpha_0)+I \left.\frac{\textrm{d}V_{T,SMR}}{\textrm{d}I} \right|_{\alpha_0}
\end{equation}
where $\alpha_0$ gives the equilibrium direction of M around which it is oscillating (assuming it to be equal to the applied field direction, as concluded from the measured 1$^\textrm{st}$ harmonic SMR response). Calculating the derivative in Eq.(\ref{eq:1storder}), using Eq.(\ref{eq:SMR}), neglecting $m_z$ and keeping in mind that $\alpha$ is dependent on $I$, gives
\begin{equation}
	\label{eq:SMRder}
	\left.\frac{\textrm{d}V_{T,SMR}}{\textrm{d}I}\right|_{\alpha_0}=\Delta R_1\sin(\alpha_0)\cos(\alpha_0)+\Delta R_1I\cos(2\alpha_0)\frac{\textrm{d}\alpha}{\textrm{d}I}
\end{equation}
The first term on the right side of Eq.(\ref{eq:SMRder}) describes the 1$^\textrm{st}$ order response (linear with $I$), showing the expected transverse SMR behavior (Eq.(\ref{eq:SMR})). The second term is a 2$^\textrm{nd}$ order response ($R_{2,SMR}$) and will therefore show up in addition to the expected SSE signal. $\frac{\textrm{d}\alpha}{\textrm{d}I}$ is the term which includes the deviation of M due to current-induced magnetic fields, and its magnitude is dependent on both $\alpha_0$ and the magnitude of the total magnetic field (applied field, coercive field and the current-induced fields). For large applied magnetic fields, the current-induced magnetic fields will have a negligible effect on M, so $\frac{\textrm{d}\alpha}{\textrm{d}I}$ goes to zero, leaving only the SSE signal in the 2$^\textrm{nd}$ harmonic signal (as is observed). Note that also in the described external heating experiments Oersted fields are generated, influencing M, but there the dynamic SMR signal will not be detected, as the SMR itself is not present.

To find an expression for $\frac{\textrm{d}\alpha}{\textrm{d}I}$, first the direction of M (given by $\alpha$) is defined, taking into account the Oersted fields (causing $\Delta\alpha$):
\begin{equation}
	\label{eq:Mdir}
	\alpha=\alpha_0+\Delta\alpha=\alpha_0+\textrm{atan}\left(\frac{B_{oe}}{B_{ex}}\right)\cos(\alpha_0)
\end{equation}
where $B_{ex}$ is the applied magnetic field and $\textrm{atan}(\frac{B_{oe}}{B_{ex}})$ is the maximum deviation of M from $\alpha_0$, which is the case for $\alpha_0=0^\circ$ ($B_{ex}$ perpendicular to $B_{oe}$, neglecting any other field contributions). From Eq.(\ref{eq:Mdir}) now $\frac{\textrm{d}\alpha}{\textrm{d}I}$ ($=\frac{\textrm{d}\alpha}{\textrm{d}B_{oe}}\frac{\textrm{d}B_{oe}}{\textrm{d}I}$) can be derived, finding
\begin{equation}
	\label{eq:dalpha}
	\frac{\textrm{d}\alpha}{\textrm{d}I}=\frac{\mu_0}{2w}\frac{B_{ex}}{B_{ex}^2+B_{oe}^2}\cos(\alpha_0)
\end{equation}
Substituting the derived equation for $\frac{\textrm{d}\alpha}{\textrm{d}I}$ in Eq.(\ref{eq:SMRder}), we calculate the expected 2$^\textrm{nd}$ harmonic SMR signal due to dynamic behavior of M, caused by the current-induced Oersted field as:
\begin{equation}
	\label{eq:R2ndHarm}
	R_{2,SMR}=\Delta R_1\frac{\mu_0}{2w}\frac{B_{ex}}{B_{ex}^2+B_{oe}^2}\cos(2\alpha_0)\cos(\alpha_0)
\end{equation}

Additional to $R_{2,SMR}$, also the SSE will be present as a 2$^\textrm{nd}$ harmonic signal, showing $\cos(\alpha_0)$ behavior with an amplitude independent from the applied magnetic field strength. The amplitude of the SSE signal can be derived from the high-field measurements shown in Fig. \ref{fig:Fig3}(b) and Fig. \ref{fig:Fig4}(b). 

\begin{figure}
\includegraphics[width=8.5cm]{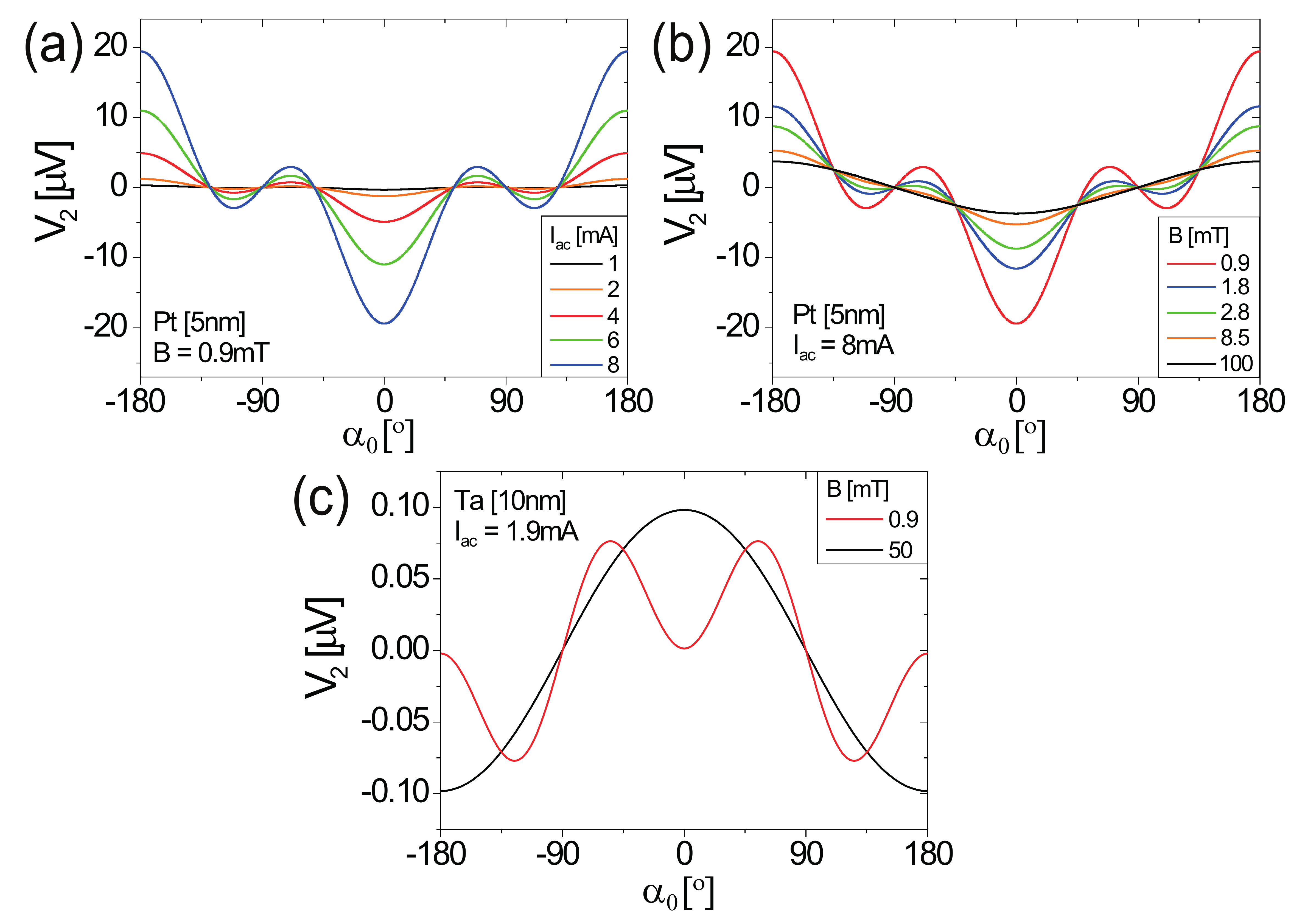}% Here is how to import EPS art
\caption{\label{fig:Fig5} 
(a) Current dependence and (b) magnetic field dependence of the calculated 2$^\textrm{nd}$ harmonic response, including the dynamic SMR (from Eq.(\ref{eq:R2ndHarm})) and the SSE for the YIG/Pt system. (c) Calculated 2$^\textrm{nd}$ harmonic response for the YIG/Ta sample using a scaling factor for $\frac{\textrm{d}\alpha}{\textrm{d}I}$ of 0.5. For all calculations the SSE signal is extracted from the measurements. 
}
\end{figure}

Figures \ref{fig:Fig5}(a) and (b) show the total calculated 2$^\textrm{nd}$ harmonic voltage signal for the YIG/Pt sample, taking into account both the SSE, extracted from the measurements, and the dynamic SMR term, given by Eq.(\ref{eq:R2ndHarm}) (where $V_{2,SMR}=(I^2/\sqrt{2})R_{2,SMR}$). For the calculation of $\Delta R_1$ system parameters from ref.\citep{VlietstraSMR2} are used. Both the calculated current dependence (Fig. \ref{fig:Fig5}(a)) as well as the calculated magnetic field dependence (Fig. \ref{fig:Fig5}(b)) show similar features as the measurements, only its magnitude and exact shape do not fully coincide. Following the explanation of the lock-in detection method as described in the appendix, we find that these discrepancies are mainly caused by the presence of a non-negligible 4$^\textrm{th}$ harmonic signal (and possibly even higher harmonics). When determining the 2$^\textrm{nd}$ order response of the system, taking into account the measured 2$^\textrm{nd}$ and 4$^\textrm{th}$ harmonic signals, the peaks observed around $\alpha_0=\pm90^\circ$ get wider and smoother, more closely reproducing the calculated signals as presented in Fig. \ref{fig:Fig5}. 

The amplitude of the overall calculated signal is slightly larger than the measurements, up to a factor 1.6 (even after taking into account the 4$^\textrm{th}$ harmonic signal). One reason for this discrepancy might be that, for the calculations, only the applied magnetic field and the generated Oersted field are taken into account, neglecting any other present fields. For example, the coercive field of the YIG is assumed to be absent, as well as the effect of spin-torque and the presence of non-uniform magnetic fields. These additional fields will influence the amplitude of the oscillations of M, giving a different value for $\frac{\textrm{d}\alpha}{\textrm{d}I}$ than the assumed deviation from only $B_{oe}$ and $B_{ex}$. Furthermore, the assumed perfect $\cos(\alpha_0)$ behavior of $\frac{d\alpha}{dI}$ might also be disturbed by the presence of these other fields. Secondly, in the calculations it is assumed that M is fully aligned with the total magnetic field, which might not always be the case, as we investigate the system applying magnetic fields approaching the coercive field of the YIG. Therefore, the definition of $\alpha_0$ can be slightly off from the assumed ideal case. Full characterization of the magnetization dynamics of the system and the magnetic field distribution would be needed to be able to give a more complete theoretical analysis of the observed features.

The same calculations have been repeated for the YIG/Ta system. The system parameters needed to calculate $\Delta R_1$ are taken from %the external heating experiments, as given in the assosciated section.
ref.\citep{JamalTaPt}, as given in section IV A. As a check, these system parameters firstly were used to calculate the 1$^\textrm{st}$ order SMR signal, finding good agreement with the measured signals (not shown). For the calculated 2$^\textrm{nd}$ harmonic signal again it is found that the amplitude of $\frac{\textrm{d}\alpha}{\textrm{d}I}$ has to be lowered to be able to reproduce the measured behavior. When lowering the calculated $\frac{\textrm{d}\alpha}{\textrm{d}I}$ by a chosen scaling factor of 0.5, results as shown in Fig. \ref{fig:Fig5}(c) are obtained. Comparing the calculated angular dependence to the measurement as shown in Fig. \ref{fig:Fig4}(b), good agreement is found in the observed behavior. Note that for the YIG/Ta system the contribution of the 4$^\textrm{th}$ harmonic term is much less pronounced, as the applied current is only 1.9mA (compared to 8mA for the YIG/Pt experiments). %However, as mentioned, to obtain these results, the amplitude of the calculated $\frac{d\alpha}{dI}$ had to be lowered.

Following the derivation of the dynamic SMR as explained above, also dynamic SSE signals can be expected. As the SSE is a 2$^\textrm{nd}$ order effect, any dynamic SSE signals are expected to appear as a 3$^\textrm{rd}$ order signal. Measurements of the 3$^\textrm{rd}$ harmonic signal indeed show additionally appearing signals at low applied magnetic fields, but these additional signals are one order of magnitude too large to be explained by the derived possible dynamic SSE signals. This shows that other higher harmonic effects are present, which makes it at this moment impossible to exclusively extract any contribution of possibly present dynamic SSE signals. 

Concluding this section, the dynamic SMR is a good candidate for explaining the observed low-field 2$^\textrm{nd}$ harmonic behavior. For both the YIG/Pt and YIG/Ta sample, the features observed in the experiments can be well reproduced by the dynamic SMR model. However, one has to keep in mind that more non-linear effects are present, such that higher harmonic signals need to be taken into account. Furthermore, the derived model is not sufficient to fully be able to reproduce the measured data. Further analysis of the magnetization dynamics in the YIG at low applied fields and high applied currents is necessary to be able to derive a more complete model. %Furthermore, we calculated that Oersted fields are non-negligible in the investigated high current and low field ranges, and they can play an important role in the magnetization dynamics of the system.

% Summary/Conclusion
\section{Summary}
We have shown the detection of the SSE in YIG/Pt and YIG/Ta samples by both external heating and current-induced heating. The external heating experiments directly show the SSE and clearly show the effect of the opposite spin-Hall angle for Ta compared to Pt. For the current-induced measurements, besides the SSE, the SMR is also present. By using a lock-in detection technique we are able to simultaneously, but separately, measure the SSE and SMR signals. Investigation of the low-field behavior of the SMR and SSE, reveals an additional 2$^\textrm{nd}$ harmonic signal. This additional signal is explained by the presence of a dynamic SMR signal, caused by alternating Oersted fields. Calculations show reproducibility of the observed 2$^\textrm{nd}$ harmonic features, however further analysis of the magnetization dynamics in the YIG is needed to derive a more complete model of the system behavior.

%------------------------------------------------------------------------------------------------------------------------------------------------------------------
%------------------------------------------------------------------------------------------------------------------------------------------------------------------
%------------------------------------------------------------------------------------------------------------------------------------------------------------------

%Acknowledgements
\section*{Acknowledgements}
We would like to acknowledge M. de Roosz, H. Adema and J. G. Holstein for technical assistance. This work is supported by NanoNextNL, a micro and nanotechnology consortium of the Government of the Netherlands and 130 partners, by the Marie Curie Actions (Grant 256470-ITAMOSCINOM), the Basque Government (PhD fellowship BFI-2011-106), by NanoLab NL and by the Zernike Institute for Advanced Materials (Dieptestrategie program).

%\nocite{*}
\bibliography{YIGPt}% Produces the bibliography via BibTeX.

\section*{Appendix}
\appendix
\subsection{Lock-in detection}
All measurements shown in the main text are performed using a lock-in detection technique. By this technique, 1$^\textrm{st}$, 2$^\textrm{nd}$ and higher order responses of a system on an applied AC-current can be determined. In general, any generated voltage can be written as the sum of 1$^\textrm{st}$, 2$^\textrm{nd}$ and higher order current-responses as:  
\begin{equation}
	\label{eq:Lockin1}
	V(t) = R_1I(t)+R_2I^2(t)+R_3I^3(t)+R_4I^4(t)+...
\end{equation}
where R$_n$ is the n$^\textrm{th}$ order response of the measured system to an applied current $I(t)$. By applying an AC-current $I(t)=\sqrt2I_0\textrm{sin}(\omega t)$, with angular frequency $\omega$ and rms value $I_0$, a lock-in amplifier can be used to detect individual harmonic voltage responses of the investigated system, making use of the orthogonality of sinusoidal functions. To extract the separate harmonic responses, the output signal and the reference input signal (a sine wave function) are multiplied and integrated over a set time. When both signals have different frequencies, the integration over many periods will result in zero signal, whereas integration of two sine wave functions with the same frequency and no phase shift will result in a non-zero signal. Besides being able to separately extract the different harmonic responses of the system, the lock-in detection technique also reduces the noise in the signal, compared to dc voltage measurements, as the measurement is only sensitive to a very narrow frequency spectrum.

The detected $n$-th harmonic signal of a lock-in amplifier at a set phase $\phi$  is defined as 
\begin{equation}
	\label{eq:Lockin2}
	V_{n}(t) = \frac{\sqrt2}{T} \int\limits_{t-T}^{t} \sin(n\omega s + \phi)V_{in}(s)ds
\end{equation} 
By evaluating Eq.(\ref{eq:Lockin2}) for a given input voltage V$_{in}$, one can obtain the different harmonic voltage signals that can be measured by the lock-in amplifier ($V_{n}$). Assuming a voltage response up till the 4$^\textrm{th}$ order, the following lock-in voltages are calculated:
\begin{equation}
	\label{eq:V1}
	V_{1} = R_1 I_0  + \frac{3}{2} R_3 I_0^3
	\;\;\;\;\;\; \textrm{for} \; \phi=0^\circ
\end{equation} 
\begin{equation}
	\label{eq:V2}
	V_{2} = \frac{1}{\sqrt2} ( R_2 I_0^2 + 2 R_4 I_0^4) \;\;\;\;\;\; \textrm{for}\; \phi=-90^\circ
\end{equation} 
\begin{equation}
	\label{eq:V3}
	V_{3} = -\frac{1}{2} R_3 I_0^3 
	\;\;\;\;\;\; \textrm{for} \; \phi=0^\circ
\end{equation} 
\begin{equation}
	\label{eq:V4}
	V_{4} = -\frac{1}{2\sqrt2} R_4 I_0^4
	\;\;\;\;\;\; \textrm{for} \; \phi=-90^\circ
\end{equation} 
So, using different lock-in amplifiers to measure the 1$^\textrm{st}$, 2$^\textrm{nd}$, 3$^\textrm{rd}$ and 4$^\textrm{th}$ harmonic voltage responses, $R_n$ can be deduced from Eqs.(\ref{eq:V1})-(\ref{eq:V4}). To detect the 2$^\textrm{nd}$ and 4$^\textrm{th}$ harmonic response, the phase of the lock-in amplifier should be set to $\phi=-90^\circ$.

Note that $V_1$ ($V_2$) does not purely scale linearly (quadratically) with $I_0$. A 3$^\textrm{rd}$ (4$^\textrm{th}$) order current dependence is also present in the measured voltage response. Thus, to obtain the 1$^\textrm{st}$ order response ($R_1$) of the measured system, not only the measured 1$^\textrm{st}$ harmonic signal ($V_1$) has to be taken into account, also the 3$^\textrm{rd}$ harmonic signal ($V_3$) has to be included:
\begin{equation}
	\label{eq:R1}
	R_1 = \frac{1}{I_0}(V_{1} + 3V_{3})
\end{equation} 
Similarly, the 2$^\textrm{nd}$ order response is calculated as
\begin{equation}
	\label{eq:R2}
	R_2 = \frac{\sqrt2}{I_0^2}(V_{2} + 4V_{4})
\end{equation}

\begin{figure}
\includegraphics[width=8.5cm]{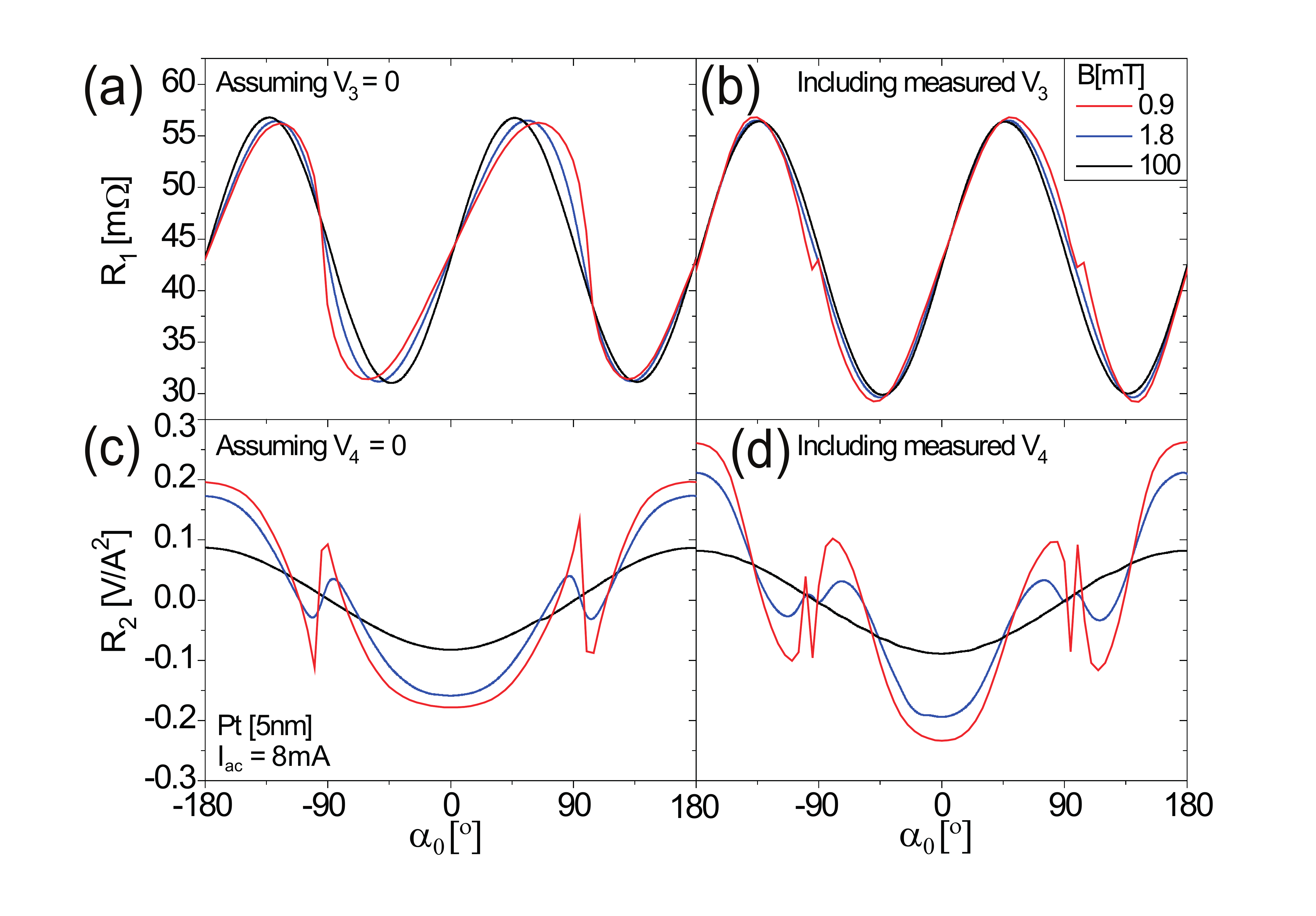}% Here is how to import EPS art
\caption{\label{fig:Fig6} 
Evaluation of Eq.(\ref{eq:R1}) for a selected set of measurements: (a) assuming $V_3=0$ and (b) including the measured values of $V_3$. Similarly for Eq.(\ref{eq:R2}): (c) assuming $V_4=0$ and (d) including the measured values of $V_4$. 
}
\end{figure}

For the current-induced SSE and SMR measurements described in the main text, Fig. \ref{fig:Fig6} shows the effect of including the higher harmonic responses of the system (up to the 4$^\textrm{th}$ harmonic), compared to assuming them to be negligible. For this comparison Eq.(\ref{eq:R1}) and Eq.(\ref{eq:R2}) are used. In Fig. \ref{fig:Fig6}(a) and (c) $V_3$ and $V_4$ are assumed to be zero, whereas in Fig. \ref{fig:Fig6}(b) and (d) the measured 3$^\textrm{rd}$ and 4$^\textrm{th}$ harmonic response have been taken into account, respectively. From these figures it can be concluded that in our low-field measurements the higher harmonic signals are not negligibly small, and thus should be accounted for. 

In Fig. \ref{fig:Fig2}(a) and (b) of the main text, the 1$^\textrm{st}$ order response of the system is plotted, taking into account the measured 3$^\textrm{rd}$ harmonic voltage response as derived in Eq.(\ref{eq:R1}) and shown in Fig. \ref{fig:Fig6}(b). For the other figures, regarding the 2$^\textrm{nd}$ order response of the system, the measured lock-in voltage is directly plotted ($V_2$). To find the 2$^\textrm{nd}$ order response of the system Eq.(\ref{eq:R2}) should be evaluated, including both $V_2$ and $V_4$. Taking this correction into account, the observed features of $V_2$ slightly change, as is shown in Fig. \ref{fig:Fig6}(d), more closely following the expected behavior from the calculated dynamic SMR signal. 

%\begin{figure*}
%\includegraphics[width=17cm]{FigDC}% Here is how to import EPS art
%\caption{\label{fig:FigDC} 
%(a) DC-current dependence of the 1$^\textrm{st}$ harmonic voltage signal for applied DC-currents of $\pm$6mA and 0mA (inset), having an additional AC-current of 100$\mu$A. (b) Average of the measured curves for I$_DC=\pm$6mA, revealing the SMR signal. (c) Angular dependence of extracted SMR signal for applied fields of 50mT and 0.9mT. (d) Difference of measured curves for I$_DC=\pm$6mA, revealing the SSE signal. (e) Angular dependence of extracted SSE signal for applied fields of 50mT and 0.9mT. Red dashed lines in figures (b,c) and (d,e) mark the maximum signal for the SMR and SSE, respectively, showing consistency between both measurement methods.
%}
%\end{figure*}

%\subsection{DC-current measurements}
%M. Schreier et al.\citep{SSEGoennenwein} performed all their measurements using only DC-currents. By applying a DC-current with a small modulation AC-current on top, we used the lock-in detection technique also to perform DC-like measurements, to compare our results to Schreier et al.. 

%Now, all signals appear simulanuously as a 1$^\textrm{st}$ harmonic voltage signal, and SMR and SSE are no longer directly separately measurable.

%By adding and subtracting the results for an applied DC-current of $\pm$6mA, we achieve curves which show exactly the behavior as directly observed when applying large AC-currents. 

\end{document}